\pdfoutput=1

\documentclass[11pt]{article}

\usepackage{ACL2023}

\usepackage{times}
\usepackage{latexsym}

\usepackage[T1]{fontenc}

\usepackage[utf8]{inputenc}

\usepackage{microtype}

\usepackage{graphicx}
\usepackage{booktabs}
\usepackage{multicol}
\usepackage{multirow}
\usepackage{amsmath}
\usepackage{amssymb}
\usepackage{bbding}
\usepackage{subfigure}
\usepackage{colortbl}
\usepackage{pifont}
\usepackage{bbm}
\usepackage{makecell}
\usepackage{bbding}
\usepackage{mathrsfs}
\usepackage{bm}
\usepackage{dsfont}
\usepackage{tabu}
\NewDocumentCommand\emojiowl{}{
    $\vcenter{\hbox{\includegraphics[height=1.5em]{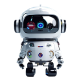}}}$
}

\newcommand{\ourmethod}{\textsc{SEvenLLM}}
\newcommand{\instruct}{\textsc{SEvenLLM}-Instruct}
\newcommand{\benchmark}{\textsc{SEvenLLM}-Bench}
\newcommand{\im}{Select-Instruct}
\title{\ourmethod{}\emojiowl{}: Benchmarking, Eliciting, and Enhancing Abilities of Large Language Models in Cyber Threat Intelligence}

\author{
  Hangyuan Ji\textsuperscript{\rm 1}, 
  Jian Yang\textsuperscript{\rm 1 \thanks{\ Corresponding Author.}},
  Linzheng Chai\textsuperscript{\rm 1}, 
  Chaoren Wei\textsuperscript{\rm 1}, 
  Liqun Yang\textsuperscript{\rm 1}, \\
  {\bf Yunlong Duan}\textsuperscript{\rm 1}, 
  {\bf Yunli Wang}\textsuperscript{\rm 1}, 
  {\bf Tianzhen Sun}\textsuperscript{\rm 1}, 
  {\bf Hongcheng Guo}\textsuperscript{\rm 1}, \\
  {\bf Tongliang Li}\textsuperscript{\rm 1},
  {\bf Changyu Ren}\textsuperscript{\rm 1},
  {\bf Zhoujun Li}\textsuperscript{\rm 1}\\
  \textsuperscript{\rm 1}State Key Laboratory of Complex \& Critical Software Environment, Beihang University \\
  \{jhy\_1,jiaya, challenging, weichaoren, lqyang, tonyliangli, cyren, lizj\}@buaa.edu.cn; \\
 \\
}

\begin{document}
\maketitle
\begin{abstract}
To address the increasing complexity and frequency of cybersecurity incidents emphasized by the recent cybersecurity threat reports with over 10 billion instances, cyber threat intelligence (CTI) plays a critical role in the modern cybersecurity landscape by offering the insights required to understand and combat the constantly evolving nature of cyber threats. Inspired by the powerful capability of large language models (LLMs) in handling complex tasks, in this paper, we introduce a framework to benchmark, elicit, and improve cybersecurity incident analysis and response abilities in \textbf{LLM}s for \textbf{S}ecurity \textbf{Event}s (called \textbf{\ourmethod{}}). Specifically, we create a high-quality bilingual instruction corpus by crawling cybersecurity raw text from cybersecurity websites to overcome the lack of effective data for information extraction. Then, we design a pipeline to auto-select tasks from the tasks pool and convert the raw text into supervised corpora comprised of question and response. The instruction dataset \instruct{} is used to train cybersecurity LLMs with the multi-task learning objective (28 well-designed tasks) for augmenting the analysis of cybersecurity events. Extensive experiments in our curated benchmark (\benchmark{}) demonstrate that \ourmethod{} performs more sophisticated threat analysis and fortifies defenses against the evolving landscape of cyber threats\footnote{\url{https://github.com/CSJianYang/SEevenLLM}}.
\end{abstract}

\section{Introduction}
In the rapidly evolving landscape of cyberspace, the intricate and diverse nature of cybersecurity postures is undergoing exponential growth in complexity \cite{graph_cti,ml_security,alam2022cybersecurity}. The digital realm is witnessing an unprecedented surge in cybersecurity incidents, with over 10 billion events impacting networks globally, which presents an immense challenge for cybersecurity professionals and analysts. 

\begin{figure}[t]
\centering
\includegraphics[width=1.0\linewidth]{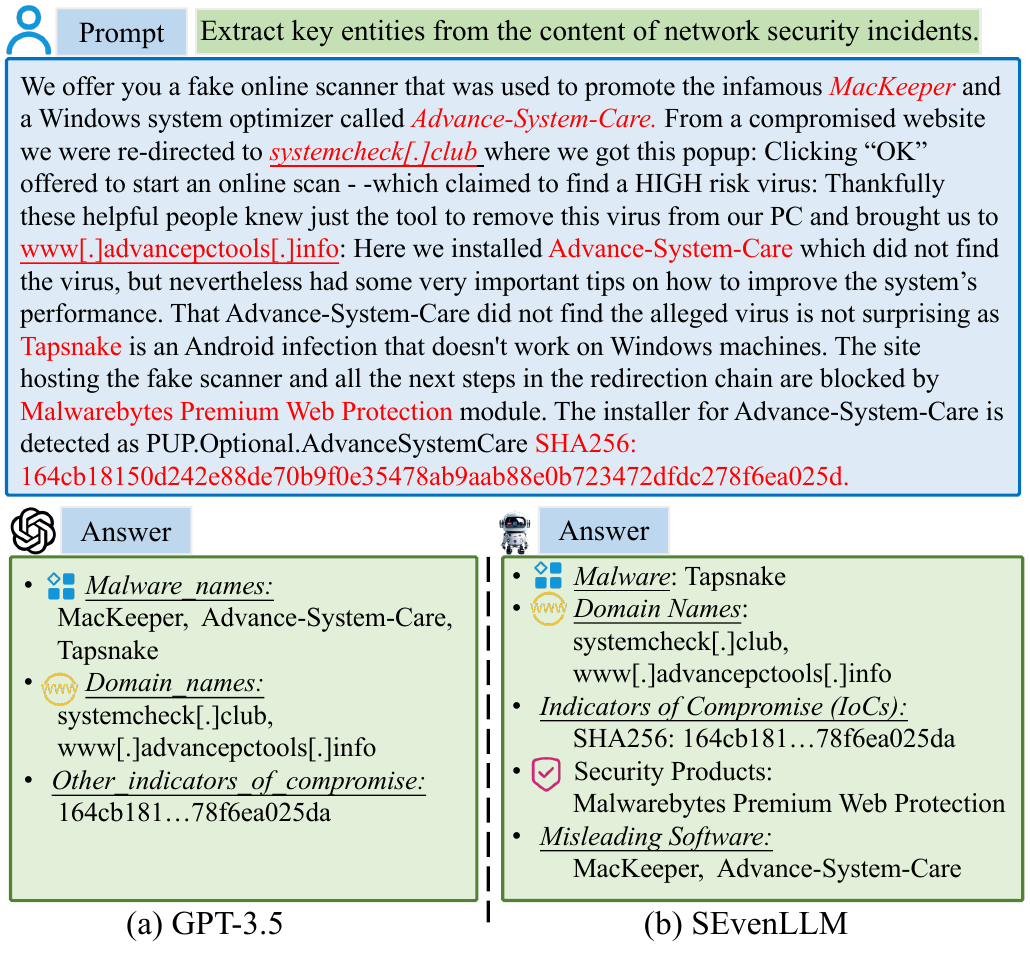}
\caption{Comparison between GPT-3.5 and our proposed model \ourmethod{}. }
\vspace{-10pt}
\label{intro}
\end{figure}

Cyber threat intelligence (CTI) \cite{kg_cti,cybersecurity_sota,arazzi2023nlp,time_cti} is proposed to understand and anticipate cyber threats for cybersecurity defense. However, traditional approaches often lack the requisite background knowledge needed for effective investigation. Large language models (LLMs) \cite{llama2,gpt3,gpt4} present a groundbreaking shift in the capabilities of understanding and generating language-based content. Domain-specific LLMs by instruction tuning further improve the specific field, such as code and math. For CTI, there is still no comprehensive work to encompass all required parts comprised of instruction corpora, domain-specific LLMs, and the evaluation benchmark.

In this paper, we introduce a framework to benchmark, elicit, and improve cybersecurity incident analysis and response abilities in \textbf{LLM}s for \textbf{S}ecurity \textbf{Event}s (called \textbf{\ourmethod{}}). In Figure \ref{intro}, the fine-tuned \ourmethod{} with fewer parameters can get a more complete and professional response compared to the strong common LLM GPT-3.5, emphasizing the importance of domain-specific LLM for CTI.
Specifically, we address the scarcity of high-quality, task-specific datasets for security event analysis by curating an extensive bilingual (English and Chinese) corpus of cybersecurity incident reports. To construct the supervised question-answer pair, we first use the LLM (GPT-4) to generate the candidate tasks, and then human experts correct the tasks based on threat intelligence analysis criteria and focal points proposed by security organizations and institutions, such as MITRE\footnote{\url{https://attack.mitre.org/}} and OASIS CTI TC\footnote{\url{https://oasis-open.github.io/cti-documentation/}}, to form a task pool. Given the cybersecurity raw text, we prompt the LLM to select the proper task from the task pool and then simultaneously generate the query and corresponding response of the chosen task, where we use \textbf{\im{}} to generate the instruction corpus \instruct{}. Open-source LLMs tailored for cyber threat intelligence based on different base models are fine-tuned on the \instruct{} with multi-task learning. To effectively evaluate \ourmethod{} for cyber threat intelligence, we construct an evaluation benchmark \benchmark{} to comprehensively evaluate the performance of LLMs for CTI, thereby bridging the gap between the common field and the cybersecurity field in understanding and generation.

The main contributions are summarized as:

\begin{itemize}
  \item  We create a high-quality bilingual multi-task instructional corpora \instruct{} by crawling cybersecurity texts and adopting the dataset construction method \im{} to address the data scarcity of CTI, where the dataset ensures the practical utility and relevance of the dataset in real-world scenarios.

  \item Based on the open-source base LLMs, \ourmethod{} tailored for cyber threat intelligence is fine-tuned on \instruct{} to assist in the automated and intelligent processing of security incidents. \ourmethod{} aims to streamline the analysis process and reduce the reliance on human expertise, thus accelerating and enhancing the capabilities of analysts in threat identification and response.

  \item To fill in the gaps in the evaluation for cyber threat intelligence, we construct an evaluation benchmark \benchmark{} comprised of multiple-choice questions and query-answer questions for assessing the performance of LLMs within the context of cybersecurity.
\end{itemize}

\begin{figure*}[h!]
\begin{center}
	\includegraphics[width=0.95\textwidth]{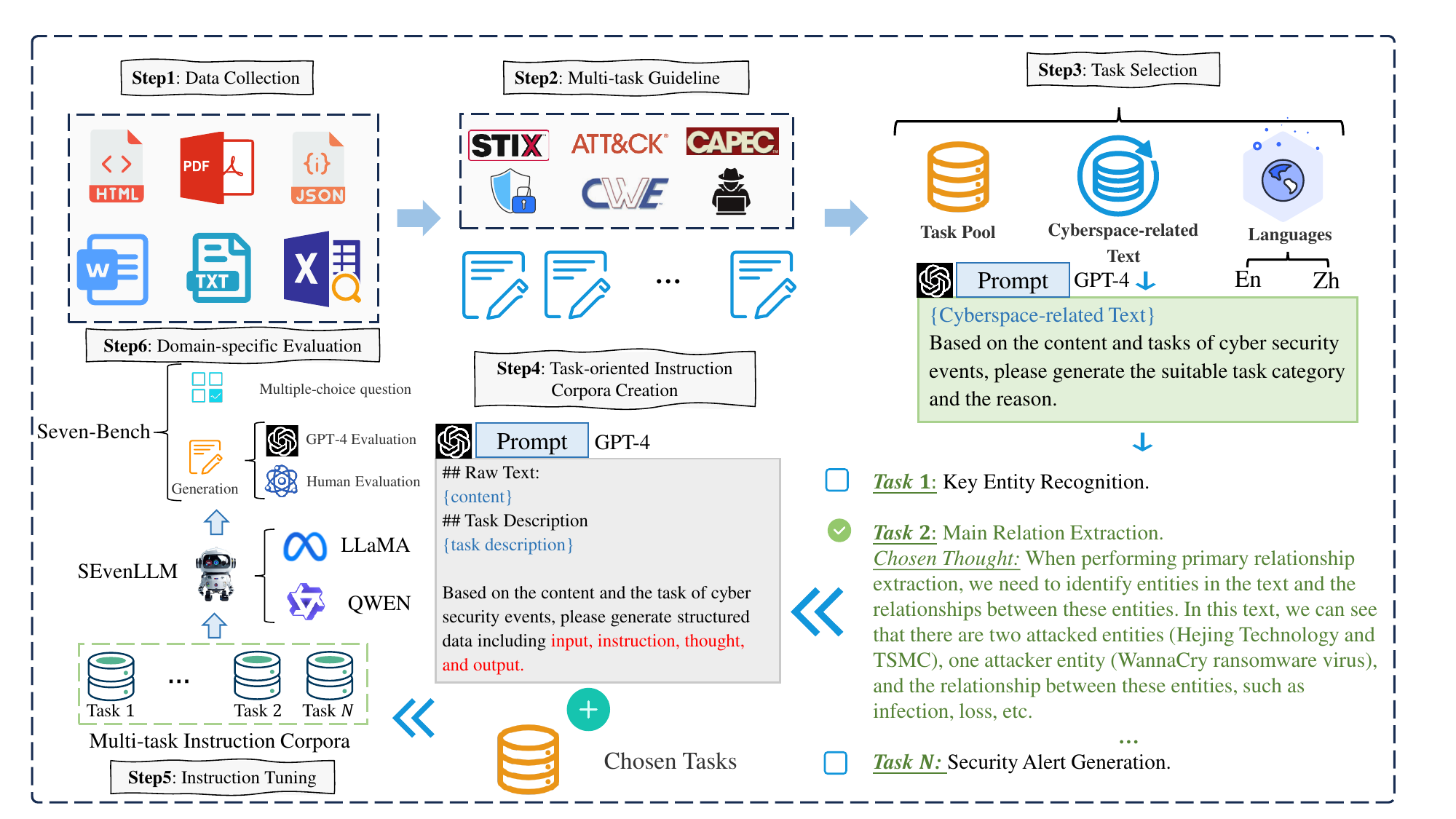}
        \caption{Overview of \ourmethod{}. By crawling different formats of files from the Internet, we collect bilingual (English and Chinese) collection of cybersecurity incident reports. First, we adopt LLMs to produce potential tasks and refine them to create a task pool. Given raw cybersecurity texts, we use \im{} select a proper task and generate the query and its answer. The open-source LLMs are further fine-tuned on \instruct{} with multi-task learning objectives tailored for CTI. A curated CTI evaluation benchmark \benchmark{} is created to compare \ourmethod{} with other baselines.
    }
	\label{overview}
	\vspace{-10pt}
\end{center}
\end{figure*}

\section{\ourmethod{}}
\subsection{Model Overview}
In Figure \ref{overview}, we propose a framework (\ourmethod{}) for assessing and enhancing the capabilities of LLMs in analyzing and responding to cybersecurity incidents, specifically referred to as security events. First, we collect extensive bilingual (English and Chinese) collection of cybersecurity incident reports. To create supervised question-answer pairs, we initially employ LLMs to produce potential tasks, which are then refined by human experts to create a collection of tasks (a task pool). Using raw cybersecurity texts, we prompt the LLM to pick an appropriate task from this collection and then generate both the query and its answer for the selected task (\im{}). This process involves using \im{} to create an instruction corpus \instruct{}. The open-source LLMs (e.g. Llama and Qwen) are further fine-tuned on \instruct{} with multi-task learning objectives tailored for cyber threat intelligence (CTI). To accurately assess the effectiveness of \ourmethod{} for CTI purposes, we establish a benchmark evaluation, \benchmark{}, designed to thoroughly evaluate the performance of LLMs in CTI, which aims to close the gap in understanding and generation capabilities between the general and cybersecurity domains.

\subsection{Benchmark Construction}

\paragraph{Data Collection and Preprocessing}
To construct the instruction corpora of the cyber threat intelligence (CTI), we amass a collection of over ten thousand cybersecurity incident websites from leading domestic and international cybersecurity vendors, ensuring the relevance and breadth of the cybersecurity incidents encompassed (from 2004 to now). The collection data includes official reports from the websites of security vendors and published media news by internet companies. We duplicate the gathered reports by title and extract text-based information from the different categories of security event reports after removing low-quality data with heuristic rules. Finally, the curated collection contains $6,706$ English and $1,779$ Chinese high-quality reports as the seeds.

\paragraph{\im{}}
To improve the quality and informational value of the generated instruction-answer datasets, we have improved the self-instruct\cite{self_instruct} method specifically for the task of cybersecurity incident analysis. Our approach involves two steps. In the first step, we input the raw corpus and use carefully designed prompts to enable the LLM to select a specified number and format of tasks from a task pool, producing task seeds. In the second step, we input both the task seeds and the raw corpus, requiring the LLM to extract valuable portions of the corpus based on the tasks to generate instructions and corresponding answers. During the generation of instructions and responses, we also require the LLM to provide its reasoning process and justification\cite{zeroshot_cot}.
 Testing results demonstrate that, using our method, out of 44,240 generated data points, there were only 24 erroneous data points and 37 empty outputs.  In contrast, the self-instruct method, when used to generate instructions and answers for all tasks, produced 41,296 data points, of which 78 were erroneous and 1,182 were empty outputs.

\paragraph{\instruct{}}
The analysis of cybersecurity incidents is categorized into understanding and generation tasks, encompassing 28 subcategories in Figure \ref{tab:multitask}. The \textit{understanding task} in cybersecurity incident analysis involves transforming unstructured data into structured data by extracting entities and relationships from cybersecurity incidents, aiming at acquiring well-defined knowledge such as identifying advanced persistent threat (APT) groups, attack techniques, and the relationships between APT groups and techniques. For cybersecurity professionals, the \textit{generation tasks} (e.g. generating summaries and suggesting mitigation measures) provide straightforward knowledge for the general analyst and are used for data fusion and information compression of cybersecurity incident content. In Figure \ref{tab:multitask}, we randomly select the raw text feed into the LLM to generate the task name and then correct them with the human experts to design 13 understanding tasks and 15 generation tasks. Given the cybersecurity-related raw text, we use \im{} method to generate the supervised instruction corpora \instruct{}.

\paragraph{\benchmark{}}
We use multiple-choice questions and query-answer questions to evaluate the performance of \ourmethod{} in CTI. In Table \ref{benchmark_statistics}, \instruct{} contains nearly 90K samples and \benchmark{} has 1300 test samples. Especially, for the multiple-choice question (MCQ), $364/1$ denotes that the length of the prompt is $364$ and the length of the output is $1$ (A, B, C, or D).










\begin{table}[t]
\resizebox{1.0\columnwidth}{!}{
\begin{tabular}{l|ccccc}
\toprule
\multirow{2}{*}{}      & \multirow{2}{*}{Problem} & \multicolumn{2}{c}{Zh} & \multicolumn{2}{c}{En} \\ 
                       &                               & Size          & Len          & Size          & Len          \\ \midrule
Seed & Raw               &  1779           &      4798       &     6706        &     900        \\\midrule
\multirow{2}{*}{Train} & MCQ               &     3000        &    347/1         &       3000      &      121/1       \\
                       & QA                    &     41218        &     384/196        &     44183        &        184/90     \\ \midrule
\multirow{2}{*}{Test}  & MCQ               &       50      &       364/1      &        50     &      138/1       \\
                       & QA                    &       600      &       450/236     &     600        &  167/91  \\        
\bottomrule
\end{tabular}}
\caption{Statistics of \benchmark{}. The training data contains two types of questions: multiple-choice questions (MCQ) and  query-answer questions(QA). The supervised data derives from high-quality raw texts (Seed) collected from the Internet. Here, we calculate the length of Chinese at the token level and the length of English by splitting space tokens. }
\label{benchmark_statistics}
\vspace{-5pt}
\end{table}

\subsection{Large Language Model for CTI}
Given a cybersecurity raw text $c \in L_{all}=\{D^{r}_{L_{i}}\}_{i=1}^{m}$ ($L_{i}$ is the language of $c$ and $L_{all}$ contains Chinese and English in our work), we prompt the LLM $\mathcal{M}$ to select the proper task $T_{k}$, where $T_{k} \in T_{all}=\{T_{j}\}_{j=1}^{n}$ and the task pool $T_{all}$ contains $13$ understanding tasks and $15$ generation tasks, as shown in Table \ref{tab:multitask}. Conditioned on \textcolor{blue}{\{Pre-defined tasks\}} and \textcolor{orange}{\{Cybersecurity-related Raw Text\}}, we use the prompt ``Based on the content and tasks of cyber security events, please generate suitable task category and the reason'' in Figure \ref{overview} to generate the target task $T_{k}$ with the chosen reason. Then, we use the raw text $c$ and the chosen task $T_{k}$ to generate the query, the corresponding response, and the thought $q_{k}, a_{k},t_{k}$ to construct the supervised fine-tuned corpora $D_{all}=\{D_{i}\}_{i=1}^{m}$, where $D_{all}$ is the supervised instruction corpus containing $m$ tasks and $(q_{k}, a_{k}) \in D^{s}_{all}$. In Figure \ref{overview}, we feed the prompt ``\textcolor{orange}{\{Cybersecurity-related Raw Text\}}. \textcolor{blue}{\{Task Definition\}}. Based on the content and the task of cyber security events, please generate structured data including input, instruction, thought, and output'' into the LLM $\mathcal{M}$ to generate the supervised pair $q_{k}, a_{k},t_{k}$. The CoT generation process can be described as:
\begin{MiddleEquation}
\begin{align}
    P(a_{k},t_{k}|q_{k}) = \prod_{j=1}^{L}P(a_{k}^{j}|a_{k}^{<j};q_{k},t_{k},\mathcal{M})
    \label{problem_definition}
\end{align}
\end{MiddleEquation}where $t_{k}$ is the reasoning step \cite{cot,zeroshot_cot} for the answer $a_{k}$. $a_{k}^{j}$ is the $j$-th token of the answer $a_{k}$.

\subsection{Multitask Instruction Tuning}
Given the bilingual multitask instruction corpora $D_{all}=\{D_{i}\}_{i=1}^{m}$, where $D$ contains $m=2$ languages and $n=28$ tasks. The base LLM is jointly trained on the multitask corpus $D_{all}$:
\begin{MiddleEquation}
\begin{align}
\begin{split}
    \mathcal{L}_{all} &=-\sum_{i=1}^{m} \mathbb{E}_{q_{k}^{L_{i}},a_{k},t_{k} \in D_{i}} \left[ \log P(a_{k},t_{k}|q_{t_k}; \theta) \right] 
    \label{xsft}
\end{split}
\end{align}
\end{MiddleEquation}where $q_{k}$ is the query, $t_k$ is the thought, and $a_{k}$ is the response.

\begin{table*}[h!]
\centering
\resizebox{0.9\textwidth}{!}{
\begin{tabular}{l|l}
\toprule
Task Name  & Task Description                                                                                                 \\ \midrule
\multicolumn{2}{c}{Understanding Task}                    \\ \arrayrulecolor{lightgray}\midrule
(1) Key Entity Recognition                         & \makecell[l]{Identify the main entity information in the text, such as attacker organization, victim type, main person, \\the common vulnerabilities and exposures (CVE), email address, virtual account, IP address, and the indicators of compromise (IOCs).}        \\ 
(2) Main Relation Extraction                       & \makecell[l]{Extract the relationships between major entities such as attacker, victim, attack method and so on. Through \\ relationship extraction, connections between entities can be established to help cybersecurity \\ experts better understand the content and context of threat intelligence.} \\ 
(3) Important Event Extraction                     & \makecell[l]{Key information such as the type, time, location, and impact of the event can be identified \\ through critical event extraction.}                                \\ 
(4) Malware Feature Extraction                     & \makecell[l]{Extract the malware features mentioned in the text, such as file characteristics, means of exploitation, and harm caused.}           \\ 
(5) Cybersecurity Event Classification             & \makecell[l]{The classification results or category characteristics such as event type, severity, etc. are extracted from the security \\ event information and structurally labeled.}       \\
(6) Attack Tool Identification                     & Tools and toolchains utilized in the attack are identified and extracted.   \\
(7) Domain Intelligence Acquisition                & \makecell[l]{Domain names often involve information about phishing sites and locations, obtaining \\ the domain name used by the attacker to look for potential relevance.}    \\
(8) Time Element Acquisition                       & The timing of cybersecurity events is extracted and the timeline is scrutinized and parsed in detail.      \\
(9) Network Protocol Utilization                   & \makecell[l]{Extract the network protocols used in the attack, which may include some information \\ containing the attacker's characteristics.}           \\
(10) Enc-Dec Algorithm Identification & \makecell[l]{The process of identifying the encryption or decryption algorithm used in a text and determining the \\ specific algorithm used for the encryption or decryption operation in the text.}     \\
(11) Vulnerability Intelligence Extraction         & \makecell[l]{Identify the main information of the vulnerability described in the text, the vulnerability number, \\ and the way the vulnerability is exploited.}       \\
(12) Attacker Information Extraction                & \makecell[l]{Characterize the attacker's place of origin, ip, language, unique fingerprints during the attack, \\ and attack behavior and confrontation ideas}     \\
(13) Attack Target Intelligence Gathering           & \makecell[l]{Analyze the main characteristics of the attacked target, such as operating system, software,\\ and target industry, field, geographical location, etc.}           \\ \arrayrulecolor{lightgray}\midrule
\multicolumn{2}{c}{Generation Task}        \\ \arrayrulecolor{lightgray}\midrule \arrayrulecolor{black}
(1) Vulnerability Exploitation Analysis            & Analyze and assess the exploitation of known vulnerabilities (also known as exploits). \\
(2) Attack Means Analysis                          & Analyze the means and specific methods used in attacks during cybersecurity incidents.      \\
(3) Attack Strategy Analysis                       & Analyze the attacker's tactics, attack plan, or usual methods in a cybersecurity incident.       \\
(4) Correlation Analysis                           & Analyze the connections and correlated evidence between different threat intelligence reports and cybersecurity incidents.                         \\
(5) Attack Intent Analysis                         & Analyze the attacker's potential motivation, intent, target industry, or target area.    \\
(6) Threat Analysis                                & Analyze potential threats and possible hazards in cybersecurity incidents.         \\
(7) Risk Assessment                                & Assess the risk and level of risk posed by the security incident or such attack.       \\
(8) Impact Scope                                   & Analyze the scope and impact of security incidents.                             \\
(9) Trend Prediction
& \makecell[l]{Predicts future security trends based on historical and current data, or predicts likely development \\ outcomes based on known information.}        \\
(10) Behavioral Pattern Analysis
& \makecell[l]{Research the behavioral pattern characteristics of attackers in cyber security incidents, \\ including attack timing, target selection, etc.}         \\
(11) Protection Strategy Research                   & Research and design strategies to protect against such security incidents or attacks.     \\
(12) Incident Response Planning                     & Develop an emergency response plan to deal with such incidents.           \\
(13) Security Policy Audit                          & Audit and evaluate the applicability and effectiveness of proposed security policies.        \\
(14) Summary Generation                             & \makecell[l]{A concise and complete summary of the text needs to be generated based on the content of the long text, \\ highlighting the activities and characteristics associated with the main cybersecurity incidents.}                 \\
(15) Security Alert Generation                      & Automatically generate security alerts based on analysis of events.          \\ 
\bottomrule
\end{tabular}}
\caption{List of the understanding and generation tasks.}
\label{tab:multitask}
\end{table*}

\begin{table*}[h!]
    \centering
    \resizebox{0.9\textwidth}{!}{
    \begin{tabular}{lccccccccccc}
    \toprule
        Model & Base Model & Model Size & \makecell[c]{Understanding \\ (En)}  & \makecell[c]{Generation \\ (En)}  & \makecell[c]{Understanding \\ (Zh)} & \makecell[c]{Generation \\ (Zh)} & \makecell[c]{Understanding \\ (Avg.)} & \makecell[c]{Generation \\ (Avg.)} & Avg. \\ \midrule
        Llama2-Chat \cite{llama2}            & Llama & 7B  & 55.8 & 39.8 & 52.3 & 81.8 & 54.1 & 60.8 & 57.5 &   \\
        Llama2-Chat \cite{llama2}            & Llama & 13B  & 48.1 & 39.6 & 56.4 & 74.6 & 52.2 & 57.1 & 54.7 &   \\
        Qwen1.5-Chat \cite{qwen}           & Qwen & 7B  & 76.3 & 45.6 & 81.4 & 84.8 & 78.8 & 65.2 & 72.0 &   \\ 
        Qwen1.5-Chat \cite{qwen}           & Qwen & 14B  & 75.0 & 42.6 & 74.9 & 86.7 & 75.0 & 64.6 & 69.8 &   \\ \midrule
        \ourmethod{}           & Llama & 7B  & 78.7 & \textbf{68.9} & \textbf{84.5} & 88.4 & \textbf{81.6} & 78.7 & \textbf{80.1} &   \\
        \ourmethod{} + CoT     & Llama & 7B  & 76.9 & 67.2 & 82.9 & 88.4 & 79.9 & 77.9 & 78.9 &   \\
        \ourmethod{}           & Llama & 13B  & 77.5 & 68.2 & \textbf{84.5} & \textbf{90.0} & 81.0 & \textbf{79.1} & \textbf{80.1} &   \\
        \ourmethod{} + CoT     & Llama & 13B  & 76.9 & 67.2 & 82.9 & 88.4 & 79.9 & 77.9 & 78.9 &   \\
        \ourmethod{}           & Qwen  & 7B & \textbf{78.9} & 66.7 & 83.3 & 88.1 & 81.1 & 77.4 & 79.3 &  \\ 
        \ourmethod{}           & Qwen  & 14B & 78.6 & 66.8 & 84.4 & 88.4 & 81.5 & 77.6 & 79.6 &  \\ 
    \bottomrule 
    \end{tabular}}
    \caption{Rouge-L scores of our method and previous baselines for downstream generation tasks.}
     \label{tab:rougelscore}
\end{table*}

\begin{table*}[h!]
    \centering
    \resizebox{0.9\textwidth}{!}{
    \begin{tabular}{lccccccccccc}
    \toprule
        Model & Base Model & Model Size & \makecell[c]{Understanding \\ (En)}  & \makecell[c]{Generation \\ (En)}  & \makecell[c]{Understanding \\ (Zh)} & \makecell[c]{Generation \\ (Zh)} & \makecell[c]{Understanding \\ (Avg.)} & \makecell[c]{Generation \\ (Avg.)} & Avg. \\ \midrule
        Llama2-Chat \cite{llama2}           & Llama & 7B  & 43.3 & 71.363 & 45.4 & 84.5 & 44.4 & 77.9 & 61.2 &   \\ 
        Llama2-Chat \cite{llama2}           & Llama & 13B  & 35.9 & 73.8 & 43.8 & 81.6 & 39.9 & 77.7 & 58.8 &   \\ 
        Qwen1.5-Chat \cite{qwen}          & Qwen & 7B  & 66.6 & 77.2 & 74.9 & 86.6 & 70.8 & 81.9 & 76.4 &   \\ 
        Qwen1.5-Chat \cite{qwen}          & Qwen & 14B  & 63.4 & 75.1 & 74.4 & 87.5 & 68.9 & 81.3 & 75.1 &   \\ \midrule
        \ourmethod{}           & Llama & 7B  & \textbf{71.9} & 84.6 & 80.4 & 87.3 & \textbf{76.2} & 85.9 & 81.1 &   \\ 
        \ourmethod{} + CoT     & Llama & 7B  & 69.7 & 84.3 & 79.4 & 86.7 & 74.6 & 85.5 & 80.1 &   \\ 
        \ourmethod{}           & Llama & 13B  & 71.0 & 84.5 & \textbf{81.4} & 89.3 & \textbf{76.2} & 86.9 & \textbf{81.6} &   \\ 
        \ourmethod{} + CoT     & Llama & 13B  & 69.3 & \textbf{85.3} & 79.4 & 89.3 & 74.4 & \textbf{87.3} & 80.9 &   \\ 
        \ourmethod{}           & Qwen  & 7B & 71.5 & 82.7 & 79.4 & \textbf{89.4} & 75.5 & 86.1 & 80.8 &  \\ 
        \ourmethod{}           & Qwen  & 14B & 70.1 & 84.5 & 79.9 & 88.1 & 75.0 & 86.3 & 80.7 &\\ 
    \bottomrule 
    \end{tabular}}
    \caption{Semantic similarity score of generation task.}
     \label{tab:semanticscore}
\end{table*}

\begin{table*}[htbp!]
    \centering
    \resizebox{0.9\textwidth}{!}{
    \begin{tabular}{lccccccccccc}
    \toprule
        Model & Base Model & Model Size & \makecell[c]{Understanding \\ (En)}  & \makecell[c]{Generation \\ (En)}  & \makecell[c]{Understanding \\ (Zh)} & \makecell[c]{Generation \\ (Zh)} & \makecell[c]{Understanding \\ (Avg.)} & \makecell[c]{Generation \\ (Avg.)} & Avg. \\ \midrule
        Llama2-Chat \cite{llama2}            & Llama & 7B  & 70.4 & 78.9 & 65.2 & 67.7 & 67.8 & 73.3 & 70.6 &   \\
        Llama2-Chat \cite{llama2}            & Llama & 13B  & 69.2 & 79.9 & 69.4 & 73.3 & 69.3 & 76.6 & 73.0 &   \\
        Qwen1.5-Chat \cite{qwen}          & Qwen & 7B  & 78.2 & 79.5 & 79.5 & 77.5 & 78.9 & 78.5 & 78.7 &   \\ 
        Qwen1.5-Chat \cite{qwen}          & Qwen & 14B  & 82.1 & 79.6 & 79.7 & 77.7 & 80.9 & 78.6 & 79.8 &   \\ \midrule 
        \ourmethod{}           & Llama & 7B  & 82.1 & 79.7 & 80.9 & 77.9 & 81.5 & 78.8 & 80.2 &   \\ 
        \ourmethod{} + CoT     & Llama & 7B  & 82.3 & 79.7 & 79.6 & 77.5 & 81.0 & 78.6 & 79.8 &   \\ 
        \ourmethod{}           & Llama & 13B  & 83.7 & 80.3 & 80.9 & 78.0 & 82.3 & 79.1 & 80.7 & \\
        \ourmethod{} + CoT     & Llama & 13B  & 82.7 & 80.1 & 80.3 & 78.3 & 81.5 & 79.2 & 80.4 &   \\ 
        \ourmethod{}           & Qwen  & 7B & \textbf{84.1} & \textbf{80.7} & \textbf{81.3} & \textbf{78.7} & \textbf{82.7} & \textbf{79.7} & \textbf{81.2} &  \\ 
        \ourmethod{}           & Qwen  & 14B & 82.6 & 79.7 & 80.6 & 77.9 & 81.6 & 78.8 & 80.2 &\\ 
    \bottomrule 
    \end{tabular}}
    \caption{GPT-4 Evaluation of our method and previous baselines downstream generation tasks.}
    \label{tab:gpt4score}
\end{table*}
\begin{table*}[h!]
    \centering
    \resizebox{0.9\textwidth}{!}{
    \begin{tabular}{lccccccccccc}
    \toprule
        Model & Base Model & Model Size & \makecell[c]{Understanding \\ (En)}  & \makecell[c]{Generation \\ (En)}  & \makecell[c]{Understanding \\ (Zh)} & \makecell[c]{Generation \\ (Zh)} & \makecell[c]{Understanding \\ (Avg.)} & \makecell[c]{Generation \\ (Avg.)} & Avg. \\ \midrule
        Llama2-Chat \cite{llama2}            & Llama & 7B  & 11 & 18 & 13 & 11 & 24 & 29 & 53 &   \\
        Llama2-Chat \cite{llama2}            & Llama & 13B  & 20 & 21 & 14 & 15 & 34 & 36 & 70 &   \\ 
        Qwen1.5-Chat \cite{qwen}           & Qwen & 7B  & 19 & 23 & 25 & 24 & 44 & 47 & 91 &   \\
        Qwen1.5-Chat \cite{qwen}           & Qwen & 14B  & 22 & 24 & 22 & 24 & 44 & 48 & 92 &   \\ \midrule
        \ourmethod{}           & Llama & 7B  & 22 & 24 & \textbf{24} & \textbf{25} & \textbf{46} & 49 & 95 &   \\ 
        \ourmethod{} + CoT     & Llama & 7B  & 20 & 24 & 23 & \textbf{25} & 43 & 49 & 92 &   \\
        \ourmethod{}           & Llama & 13B  & \textbf{23} & \textbf{25} & 23 & \textbf{25} & \textbf{46} & \textbf{50} & \textbf{96} &   \\ 
        \ourmethod{} + CoT     & Llama & 13B  & 22 & \textbf{25} & \textbf{24} & \textbf{25} & \textbf{46} & \textbf{50} & \textbf{96} &   \\
        \ourmethod{}           & Qwen  & 7B & \textbf{23} & 23 & 23 & 24 & \textbf{46} & 47 & 93 &  \\ 
        \ourmethod{}           & Qwen  & 14B & 22 & \textbf{25} & \textbf{24} & \textbf{25} & \textbf{46} & \textbf{50} & \textbf{96} &\\ 
    \bottomrule 
    \end{tabular}}
    \caption{Results of multiple-choice question.}
     \label{tab:mcqscore}
\end{table*}

\section{Experiment Setting}
\subsection{Instruction Tuning}
\paragraph{Backbone Models}
Based on the English and Chinese pre-trained model, we adopt Llama-2-7B/13B\footnote{\url{https://github.com/meta-llama/llama}} and Qwen-1.5-7B/14B\footnote{\url{https://github.com/QwenLM/Qwen1.5}} as the foundation model to build \ourmethod{}.

\paragraph{Implementation Details}
To fine-tune all models with different sizes, we set epochs to 3 and batch size to 128. We use a cosine scheduler with a learning rate of 2e-5 and 3\% learning rate warmup. 

\paragraph{Dataset}
Based on based model of Llama2 and Qwen-1.5, \ourmethod{}s are trained on \instruct{} of nearly 85K samples generated by GPT-4.

\subsection{Evaluation}
\paragraph{Rouge-L Score} We use Rouge-L to evaluate the output of the LLM, which is tokenized by space in English and jieba\footnote{\url{https://github.com/fxsjy/jieba}} in Chinese.

\paragraph{Semantic Score} Using the multilingual sentence-transformers model\footnote{\url{https://huggingface.co/sentence-transformers/paraphrase-multilingual-MiniLM-L12-v2}} \cite{reimers-sentence-bert}, we map the text to a 384-dimensional dense vector space. This can be used for tasks such as clustering or semantic search, allowing for the evaluation of whether the values in a test dictionary can score from a semantic understanding perspective. This approach is more aligned with the open-ended nature of cyber security event analysis. Due to its support for various lengths and both Chinese and English languages, this model is applied to both understanding and generation tasks.

\paragraph{GPT-4 Score} To score the standard answers of the test set and the answers provided by the trained model, we set the evaluation criteria as \{very pool: 1, slightly poor: 2, neutral: 3, good: 4, very good: 5\} to compute the score.

\paragraph{Multiple-choice Question Score} We design 100 multiple-choice questions for two languages and two major types of tasks to comprehensively evaluate the model's capabilities. The output results were manually verified to ensure an effective assessment of the model's ability to handle objective questions.


\section{Main Results}
\paragraph{Rouge-L\&Semantic Evaluation} In Table \ref{tab:rougelscore} and Table \ref{tab:semanticscore}, \ourmethod{} based on the Llama2 and Qwen1.5 gets the best performance by fine-tuning on the instruction dataset \instruct{}. Further, we can observe that \ourmethod{} based on the Qwen tends to get better performance in Table \ref{tab:rougelscore}. Plus, \ourmethod{} + CoT gains no improvement, since the evaluated tasks do not require the complex reason. But the reasoning text can be provided to cybersecurity staff as a reference.

\paragraph{GPT-4 Evaluation} 
In Table \ref{tab:gpt4score}, compared to the results of models that have not been trained with \instruct{}, it is evident that training with the datasets provided in this work significantly improves performance. Additionally, it is observed that while higher-capacity models show improvements in multiple-choice questions, they can occasionally perform worse in Rouge-L and semantic similarity matching. Considering that cybersecurity data is challenging to collect and highly specialized, this underscores the importance of enhancing training materials in this field with datasets like those provided in this paper. It has also been observed that base models can perform correct analysis but still produce incorrect answers, indicating that the full potential of an AI model requires not only general training but also alignment with specialized corpora to yield more accurate outcomes.

\paragraph{Multiple-choice Question} In Table \ref{tab:mcqscore}, during a comparison of multiple-choice questions, it is observed that the Qwen performs less effectively than Llama2, particularly in scenarios requiring a degree of reasoning and where answers may be confusing. The LLMs with 13B and 14B parameters have been confirmed to perform better in multiple-choice question scenarios. This suggests that the enrichment of overall knowledge and reasoning capabilities with larger model sizes improves performance in multiple-choice questions.

\section{Analysis}
\label{analysis}
\begin{figure*}[t]
\centering
    \subfigure[Rouge-L score.]{
        \includegraphics[width=0.45\textwidth]{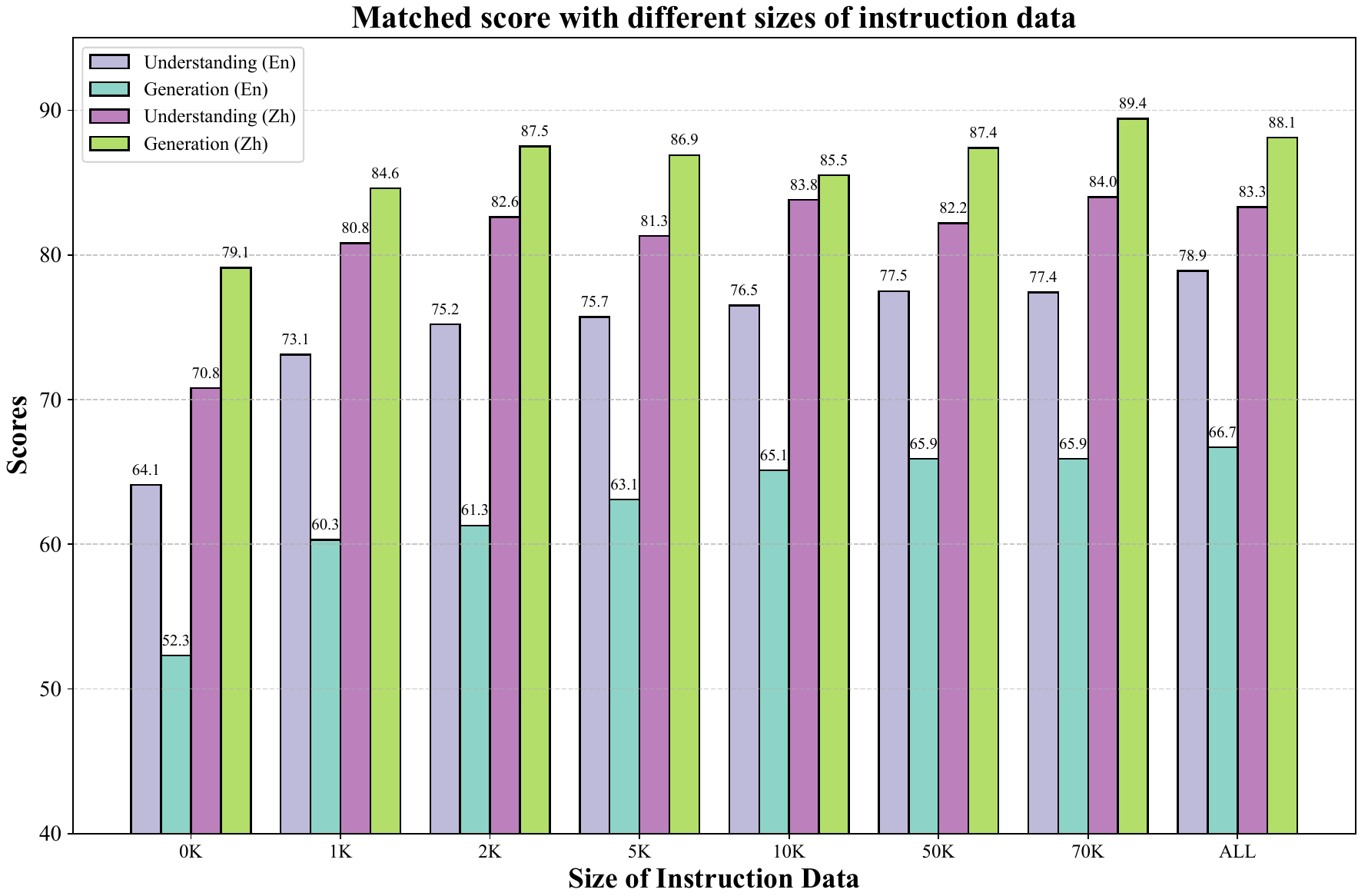}\quad
        \label{instruction_data_size_rougel}
    }
    \subfigure[Semantic Similarity score.]{
        \includegraphics[width=0.45\textwidth]{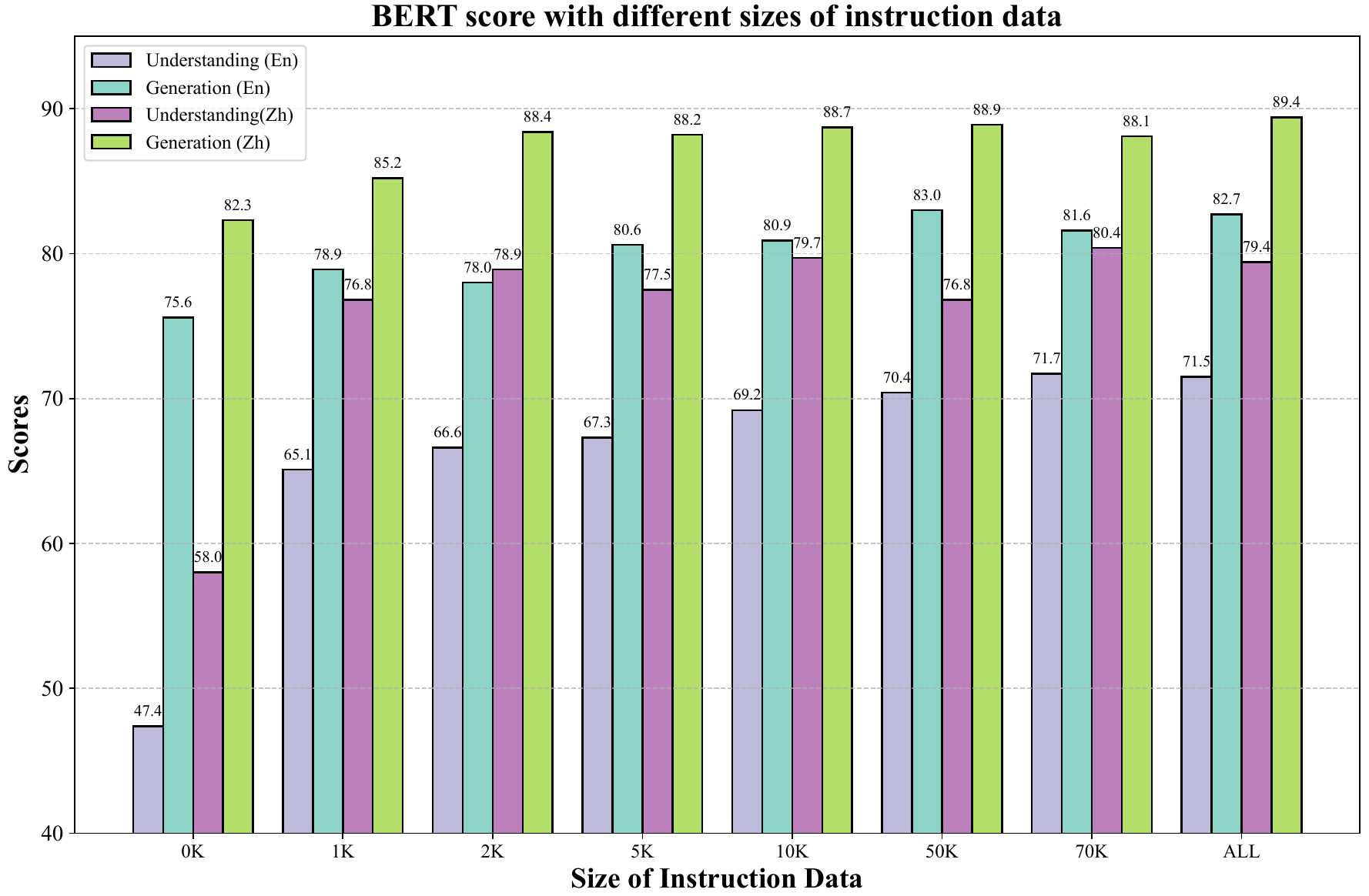}\quad
        \label{instruction_data_size_semantic_score}
    }
    \caption{Evaluation results of different instruction data sizes. }
    \label{instruction_data_size}
\end{figure*}

\paragraph{Training Data Size of Instruction Tuning}
To discuss the effect of the instruction tuning, we plot Rouge-L scores and Semantic similarity scores with different training data sizes in Figure \ref{instruction_data_size}. The performance is influenced by the size of the instruction data. We randomly sample $N=$ $\{0K, 1K, 2K, \dots, ALL\}$ sentences from the whole corpora to fine-tune the Qwen model.
With the training data size increasing, the \ourmethod{} gets better performance. Notably, only $1K$ supervised data sentences bring large improvement to the zero-shot cross-lingual NER, which benefits from knowledge transfer of the multilingual self-training. When the size of target annotated corpora is greater than 10K, our method gets exceptional performance.
From the results obtained through Rouge-L score and Semantic similarity score, the model's performance appears to reach a level close to the testing set when it scales to around the 70,000 magnitude. However, overall it appears that the larger the training dataset, the better the accuracy of the model.

\paragraph{Quality Analysis of \benchmark{}}
To ensure the quality of the \benchmark{}, we employ three experts to evaluate the quality of the test set and calculate the error rate in Table \ref{correction_benchmark}. Manual revision includes corrections for accuracy and reasonableness, inspecting the result content to ensure it meets the requirements of the instructions, is based on prior knowledge from the input, and involves removing redundant information and hallucinations, as well as adding some undetected information. This includes proofreading 100 multiple-choice questions for the correctness and uniqueness of answers.
We can that the total error rate of \benchmark{} is 17\% generated by GPT-4 and then the experts fix these errors to ensure the correctness of the test set.
\begin{table*}[!ht]
    \centering
    \resizebox{0.8\textwidth}{!}{
    \begin{tabular}{l|ccc|ccc|c}
    \toprule
        Task & $\lvert \text{Zh} \rvert$ & $\lvert \text{En} \rvert$ & $\lvert \text{All} \rvert$ & $\lvert \text{Zh (C)} \rvert$ & $\lvert \text{En (C)} \rvert$ & $\lvert \text{All (C)} \rvert$ & Ratio (\%) \\ \midrule
        Key Entity Recognition & 3 & 25 & 28 & 2 & 1 & 3 & 0.11   \\ 
        Main Relation Extraction & 0 & 0 & 0 & 0 & 0 & 0 & 0.00   \\ 
        Important Event Extraction & 12 & 28 & 40 & 2 & 2 & 4 & 0.1   \\ 
        Malware Feature Extraction  & 227 & 41 & 268 & 99 & 9 & 108 & 0.40  \\
        Cybersecurity Event Classification & 14 & 86 & 100 & 5 & 3 & 8 & 0.08  \\ 
        Attack Tool Identification & 3 & 21 & 24 & 1 & 2 & 3 & 0.13   \\ 
        Domain Intelligence Acquisition & 0 & 9 & 9 & 0 & 4 & 4 & 0.44   \\ 
        Time Element Acquisition & 2 & 36 & 38 & 1 & 0 & 1 & 0.03   \\ 
        Network Protocol Utilization & 0 & 0 & 0 & 0 & 0 & 0 & 0.00    \\ 
        Encryption-Decryption Algorithm Identification & 0 & 11 & 11 & 0 & 0 & 0 & 0 \\
        Vulnerability Information Extraction & 39 & 33 & 72 & 9 & 2 & 11 & 0.15   \\
        Attacker Information Extraction & 20 & 15 & 35 & 12 & 0 & 12 & 0.34   \\
        Attack Target Intelligence Gathering & 5 & 18 & 23 & 2 & 1 & 3 & 0.13   \\
        Vulnerability Exploitation Analysis & 0 & 0 & 0 & 0 & 0 & 0 & 0.00    \\
        Attack Means Analysis & 7 & 1 & 8 & 1 & 0 & 1 & 0.13  \\
        Attack Strategy Analysis & 14 & 3 & 17 & 2 & 0 & 2 & 0.12  \\
        Correlation Analysis & 7 & 0 & 7 & 0 & 0 & 0 & 0   \\
        Attack Intent Analysis & 2 & 1 & 3 & 0 & 0 & 0 & 0  \\
        Threat Analysis & 224 & 92 & 316 & 27 & 10 & 37 & 0.12   \\ 
        Risk Assessment & 10 & 37 & 47 & 3 & 1 & 4 & 0.09  \\ 
        Impact Scope & 8 & 20 & 28 & 0 & 2 & 2 & 0.07   \\ 
        Trend Prediction & 2 & 9 & 11 & 1 & 1 & 2 & 0.18  \\ 
        Behavioral Pattern Analysis & 0 & 0 & 0 & 0 & 0 & 0 & 0.00    \\ 
        Protection Strategy Research & 39 & 50 & 89 & 0 & 3 & 3 & 0.03   \\ 
        Incident Response Planning & 6 & 46 & 52 & 0 & 2 & 2 & 0.04  \\ 
        Security Policy Audit & 0 & 1 & 1 & 0 & 0 & 0 & 0  \\ 
        Summary Generation & 6 & 55 & 61 & 2 & 13 & 15 & 0.25   \\ 
        Security Alert Generation & 0 & 12 & 12 & 0 & 2 & 2 & 0.17  \\ \midrule
        Total & 650 & 650 & 1300 & 168 & 54 & 222 & 0.17  \\ 
        \bottomrule
    \end{tabular}}
    \caption{Correction of \instruct{}. $\lvert \text{Zh} \rvert$ denotes the number of Chinese samples in \benchmark{} while $\lvert \text{Zh} (C) \rvert$ denotes the number of samples need to be corrected.}
    \label{correction_benchmark}
    \vspace{-5pt}
\end{table*}

\paragraph{Human Evaluation}
\begin{figure}[t!]
\centering
    \subfigure[Understanding tasks.]{
        \includegraphics[width=0.43\columnwidth]{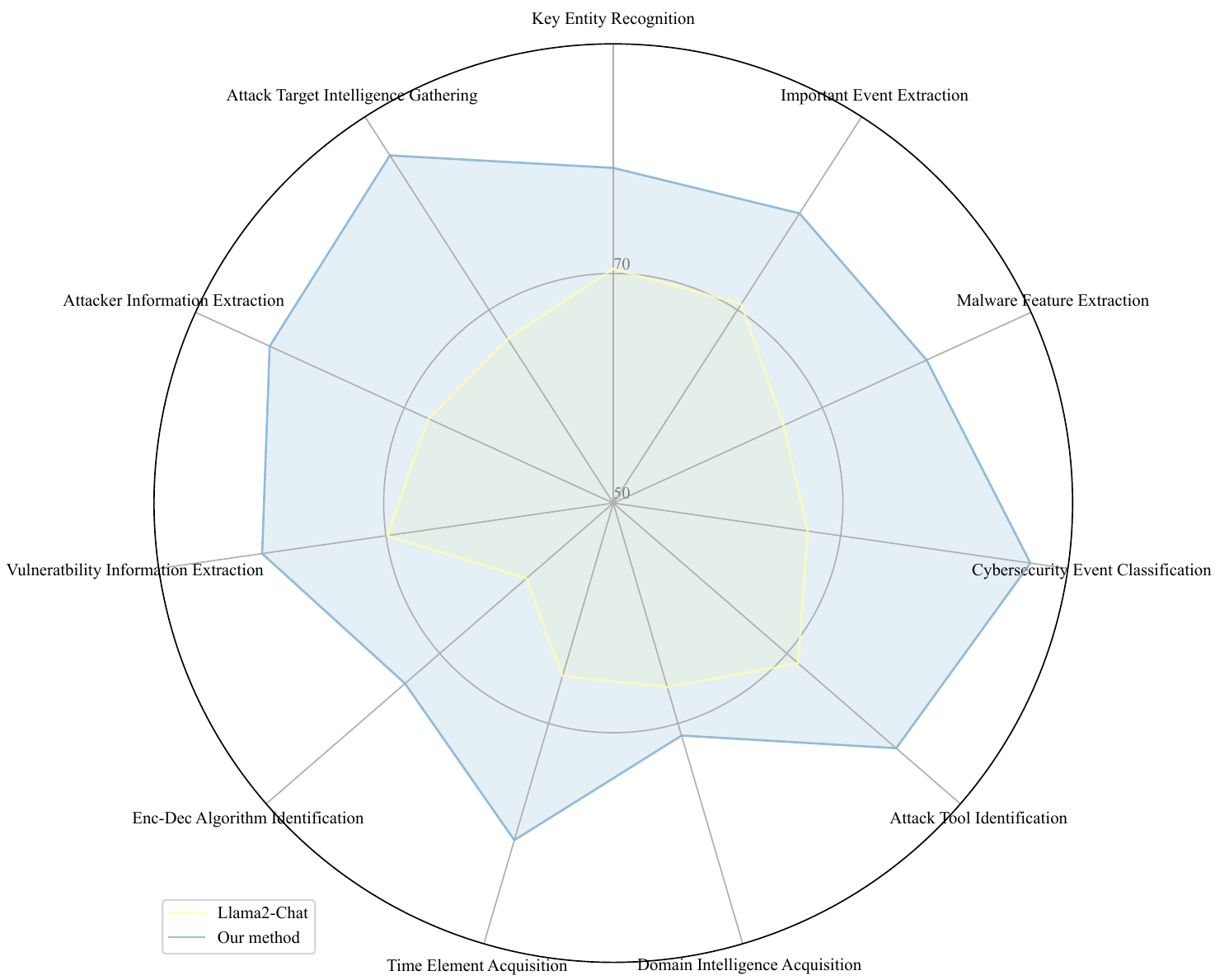}\quad
        \label{}
    }
    \subfigure[Generation tasks.]{
        \includegraphics[width=0.43\columnwidth]{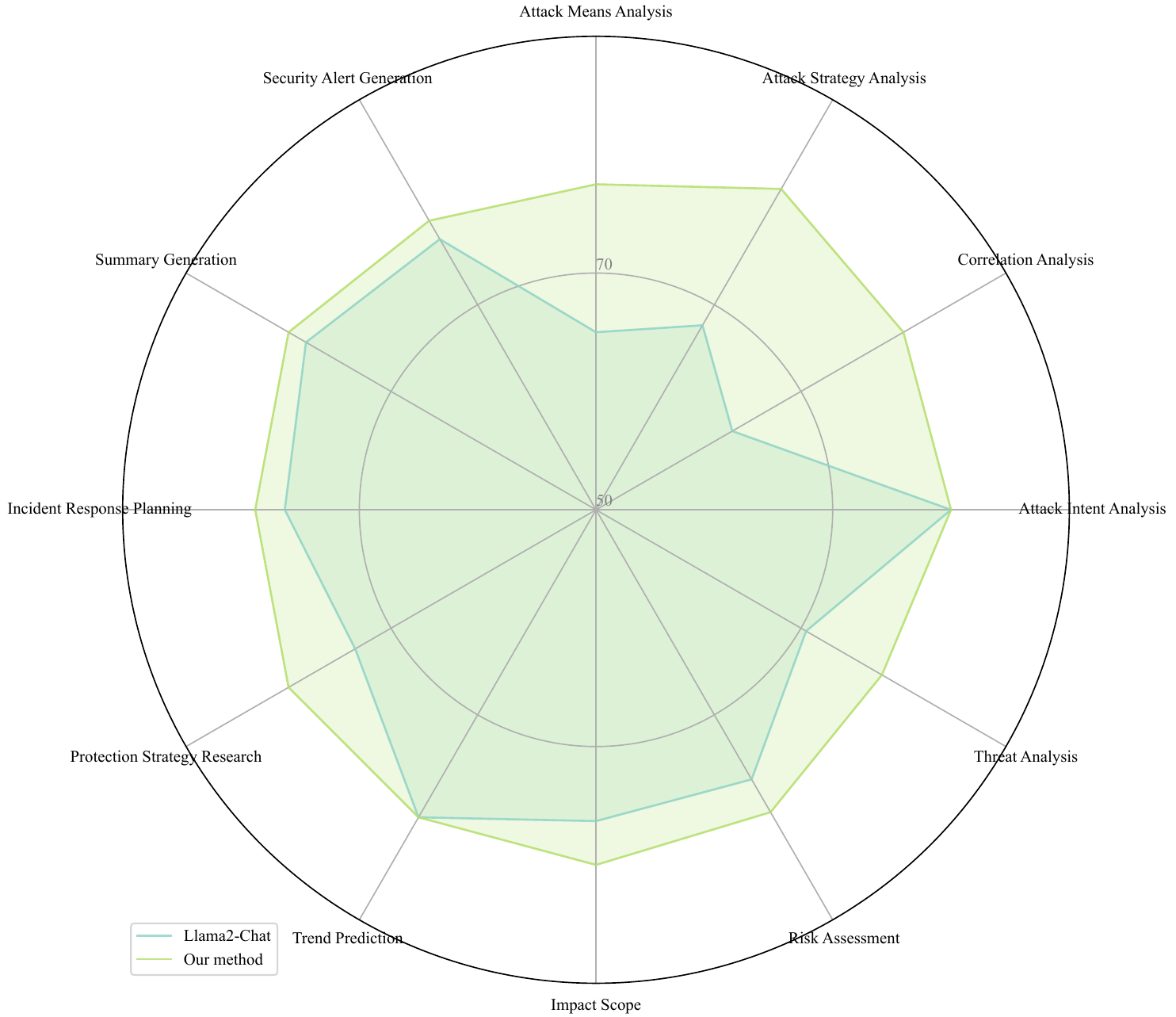}\quad
        \label{}
    }
    \caption{Comparison between \ourmethod{} with Llama-2-Chat.}
    \label{human_evaluation}
\end{figure}

To better evaluate the LLM Llama-2-Chat and \ourmethod{}, we employ five volunteers to score the test set \benchmark{} of cyber threat intelligence \{1: very poor; 2: poor; 3: neutral, 4: good; 5: very good\} based on three criteria (1) Correctness: Correctness refers to the accuracy and reliability of the information provided or the actions performed. (2) Fluency: It's about how smoothly and coherently the system can produce or interpret language, making it comprehensible and pleasant for human users. (3) Instruction Following Capability: This component assesses how effectively a system or application can comprehend and execute commands or requests given by users.

\section{Related Work}
\paragraph{Cyber Threat Intelligence} 
Cyber threat intelligence \cite{kg_cti,cybersecurity_sota,arazzi2023nlp,time_cti,camacho2024cybersecurity} has emerged as a pivotal aspect of cybersecurity practices, aimed at understanding and anticipating cyber threats for proactive defense. This field encompasses the collection, analysis, and dissemination of information regarding current or potential attacks threatening the security of information systems \cite{park2023machine,translog,guo2024lemur}. By leveraging CTI, organizations are equipped to identify specific threats and vulnerabilities, enabling them to tailor their security measures more effectively. This intelligence-gathering process draws from a wide array of sources, including historical cyber attack data, hacker forums, malware samples, and more, to provide a comprehensive view of the cyber threat landscape, which helps in enhancing the security posture against known threats and aids in predicting and mitigating future cyber attacks. Consequently, CTI plays a critical role in the strategic, operational, and tactical levels of an organization's cybersecurity framework, ensuring resilience against the evolving nature of cyber threats.


\paragraph{Domain-specific Large Language Model} 
The evolution of large language models (LLMs) \cite{llama2,qwen,gpt4} is a remarkable milestone while addressing concerns of reliability and safety. The development of the generative pre-trained Transformer (GPT) series by OpenAI progressively expands the frontiers of the field of natural language processing (NLP), 
such as named entity recognition \cite{mcl_ner,crop,c_icl}, machine translation \cite{alm,wmt2021_microsoft,soft_template,ganlm} and text style transfer~\cite{wang2019harnessing,wang2020formality,wu2020formality_dataset}.
It even has an impact~\cite{lin2023LLMRec_survey,meta2024LLMrec} on some traditional recommendation and advertising methods~\cite{wang2023arf,huang2022mbct}.
To further enhance the adaptability of LLMs to various tasks, instruction-tuning has become a commonly adopted method by utilizing instructions derived from human-curated prompts, feedback, and public benchmarks, which supports the fast development of the domain-specific LLMs, such as complex reasoning \cite{chai2024xcot,qin2024multilingual}, coding \cite{codellama,starcoder2}, IT operation \cite{guo2023owl,guo2024lemur}, and science \cite{nature_science_llm}. Instruction tuning \cite{self_instruct} is introduced to generate novel tasks and associated instructions from the ground up, offering advantages in terms of performance and cost efficiency. The breadth and variety of the instruction data correlate strongly with how well the models can generalize and handle tasks they have not encountered before.

\section{Conclusion}
In this work, we present \ourmethod{}, which marks a significant advancement in the use of LLMs for benchmarking,
eliciting, and improving cybersecurity incident analy-
sis and response abilities in cybersecurity LLMs. By leveraging a meticulously curated bilingual instruction corpus from an extensive collection of cybersecurity texts, \ourmethod{} bridges the gap in the availability of effective data for cybersecurity applications. The instruction dataset \instruct{} encompassing 28 well-conceptualized tasks is used to fine-tune \ourmethod{} based on the different foundation models (Llama and Qwen). The extensive main and analytic experiments performed on a specialized curated cybersecurity benchmark, \benchmark{}, further corroborate the efficacy of \ourmethod{} in improving the analytical capabilities and providing a robust response mechanism against cyber threats.

\section*{Limitations}
The primary data source collected focuses on English as the mainstream language for cybersecurity intelligence. Although it has been extended to include Chinese data, the analysis of cybersecurity incidents would benefit from a multilingual capability, which requires further collection and organization. In the future, we will expand \instruct{} to more languages.





\section*{Ethical Considerations}
In this study, we analyze cyber threat intelligence exclusively derived from publicly available sources provided by security companies. Although some sensitive cybersecurity terminology was used, we strictly avoided involving any personal privacy, trade secrets, or activities that could constitute a threat. Moreover, we ensured that all data handling and usage complied with ethical standards and legal regulations to maintain transparency and integrity in our research.


\bibliography{custom}
\bibliographystyle{acl_natbib}


\end{document}